\documentclass{article}

\usepackage{PRIMEarxiv}

\usepackage[utf8]{inputenc} 
\usepackage[T1]{fontenc}    
\usepackage{hyperref}       
\usepackage{url}            
\usepackage{booktabs}       
\usepackage{amsfonts}       
\usepackage{nicefrac}       
\usepackage{microtype}      
\usepackage{lipsum}
\usepackage{fancyhdr}       
\usepackage{graphicx}       
\graphicspath{{media/}}     
\usepackage{natbib}
\usepackage{amsmath}
\usepackage{threeparttable}
\usepackage{subcaption}
\usepackage{float}
\usepackage{xcolor}
\usepackage{placeins}
\usepackage{authblk}
\usepackage{appendix}
\usepackage{booktabs}

\title{Multiply-robust Estimator of Cumulative Incidence Function Difference for Right-Censored Competing Risks Data}  

\author[1]{Yifei Tian}
\author[1,*]{Ying Wu} 

\affil[1]{School of Statistics and Data Science, LPMC and KLMDASR, Nankai University, Tianjin, China}

\affil[*]{Corresponding author. Email: \texttt{ywu@nankai.edu.cn}}

\begin{document}
\maketitle

\begin{abstract}
In causal inference, estimating the average treatment effect is a central objective, and in the context of competing risks data, this effect can be quantified by the cause-specific cumulative incidence function (CIF) difference. While doubly robust estimators give a more robust way to estimate the causal effect from the observational study, they remain inconsistent if both models are misspecified. To improve the robustness, we develop a multiply robust estimator for the difference in cause-specific CIFs using right-censored competing risks data. The proposed framework integrates the pseudo-value approach, which transforms the censored, time-dependent CIF into a complete-data outcome, with the multiply robust estimation framework. By specifying multiple candidate models for both the propensity score and the outcome regression, the resulting estimator is consistent and asymptotically unbiased, provided that at least one of the multiple propensity score or outcome regression models is correctly specified. Simulation studies show our multiply robust estimator remains virtually unbiased and maintains nominal coverage rates under various model misspecification scenarios and a wide range of choices for the censoring rate. Finally, the proposed multiply robust model is illustrated using the Right Heart Catheterization dataset.
\end{abstract}


\section{Introduction}
Competing risks data are common in clinical and epidemiological studies and require specialized methodological approaches. A competing risks setting arises when an individual can experience one of several distinct types of events, and the occurrence of one event precludes the observation of other event types. Analysis of such data primarily focuses on two key measures: the cause-specific hazard, representing the instantaneous failure rate from a specific cause, and the cumulative incidence function (CIF), which quantifies the marginal probability of that cause. Conventionally, these quantities are estimated using regression frameworks such as the cause-specific Cox model or the Fine-Gray subdistribution hazard model~\cite{fine1999proportional, andersen2002competing}. To explicitly define the causal impact of an intervention, potential outcomes framework is extended to the competing risks setting~\cite{young2020causal}. This framework allows for the definition of counterfactual CIFs, thereby facilitating the estimation of the absolute risk difference between treatment arms while rigorously accounting for competing events.

Within potential outcomes framework, the primary objective is to estimate the causal effects of treatment strategies on time-to-event outcomes, typically quantified by the average treatment effect (ATE)~\cite{hirano2003efficient}. To obtain an unbiased estimation of the ATE, randomized controlled trials (RCTs) are considered the gold standard, as the random allocation of subjects ensures balance in baseline covariates across treatment groups. However, RCTs are often infeasible due to ethical constraints, high costs, or logistical complexities, which necessitates the use of observational studies. In such studies, the presence of confounders—variables that simultaneously influence treatment assignment and the outcome of interest—may lead to spurious associations rather than reflecting the true causal effect when outcomes are compared naively between treated and untreated groups. 

To address confounding, two widely used approaches are outcome regression (OR) models and inverse probability weighting (IPW). OR models directly adjust for confounders by modeling potential outcomes, whereas IPW reweights the population by the propensity score (PS)—the inverse of the treatment assignment probability—to mimic a randomized experiment~\cite{rosenbaum1983central}. However, both methods are considered singly robust, yielding unbiased estimates only when their respective models are correctly specified. This limitation led to the development of doubly robust (DR) estimators, which combine both the OR and PS models to provide consistent and asymptotically efficient estimates as long as at least one of the two component models is correctly specified~\cite{bang2005doubly, van2011targeted}. Building on this general framework, extensive researches have emerged applying doubly robust methods to survival analysis. The literature has documented several approaches to doubly robust estimation. For example, Zhang and Schaubel~\cite{zhang2012contrasting}, Bai et al.~\cite{bai2017optimal}, and Sjolander and Vansteelandt~\cite{sjolander2017doubly} defined doubly robust estimators for contrasts in expected potential failure times; Yang and Small~\cite{yang2016using} and Yang et al.~\cite{yang2020semiparametric} developed doubly robust estimators under structural accelerated failure time models; and Petersen et al.~\cite{petersen2014targeted} and Zheng et al.~\cite{zheng2016doubly} proposed targeted maximum likelihood estimators (TMLE).

Beyond these general survival applications, doubly robust methodology has been specifically adapted for the competing risks setting. Ozenne et al.~\cite{ozenne2020estimation} developed and validated a doubly robust estimator for the ATE on $t$-year absolute event risks. Lin and Trinquart~\cite{lin2022doubly} extended doubly robust methodology to the estimation of restricted mean time lost. Rava and Xu~\cite{rava2023doubly} proposed rate-doubly-robust estimators for conditional average treatment effects (CATE), enabling the use of flexible machine learning algorithms for nuisance modeling while maintaining valid statistical inference. Li et al.~\cite{li2024doubly} established a doubly robust TMLE framework for estimating CATE.  More recently, van Hage et al.~\cite{van2025doubly} introduced a doubly robust estimator constructed as a weighted combination of the IPW and OR estimators to estimate marginal cause-specific CIFs.

However, their doubly robust estimators are not infallible as they remain consistent only if at least one of the two component models is correctly specified. In complex analyses like competing risks, covariates may be numerous or their functional forms uncertain, which easily due to misspecification for both the PS and OR models. In this common failure scenario, doubly robust estimators lose their theoretical protection and yields biased results. To overcome this limitation, Han and Wang~\cite{han2013estimation} introduced the multiply robust (MR) framework based on empirical likelihood. This approach ensures consistency provided that at least one of the candidate models is correctly specified. The framework was subsequently advanced by Han~\cite{han2014further}, resolving the issue of non-unique solutions and proposing a practical computational algorithm, and by Chan and Yam~\cite{chan2014oracle}, extending multiply robust methods to the calibration of survey data with missingness. 

Building on these advances, Wang et al.~\cite{wang2023multiply} recently applied the multiply robust framework to the estimation of causal treatment effects for single-event survival outcomes. In this article, we develop a multiply robust framework tailored specifically to competing risks data. We evaluate the robustness and performance of the proposed method through extensive simulation studies and an application to a real data. 

The rest of this article is organized as follows. The following section introduces key notations, definitions, and the proposed multiply robust estimators for competing risks data, including the associated component models and pseudo-values. In the ``Simulation analysis'' section, the finite sample performances of the proposed method are evaluated. The practical use of the proposed multiply robust method is illustrated with a real-data application in the “Real data analysis” section. Finally, some remarks and discussions are presented in the ``Discussion” section.

\section{Methodology}
\subsection{Notation}
Consider a failure time study in which there exist $K$ mutually exclusive competing failure events or causes. Let $(T, \epsilon)$ denote the pair of the underlying failure time and the cause of failure, where $\epsilon\in \{1,2,\dots,K\}$ and $K<\infty$. Let $C$ be the potential censoring time. The observed event time is the minimum of the failure time and the censoring time, denoted by $\tilde{T} = \min(T,C)$. The observed failure indicator is defined as $\delta=\mathbb{I}(T\le C)$, where $\mathbb{I}(\cdot)$ is the indicator function. Then the observed cause of failure is given by $\tilde{\epsilon} = \delta \epsilon$, which equals the actual cause $\epsilon$ if an event is observed and 0 if censored. We let $\boldsymbol{X} = (X_{1}, \ldots, X_{p})^T$ denote a $p$-dimensional covariate and $A$ be the treatment indicator taking values from $\{0, 1\}$.

In the observational study, we are interested in estimating ATE of specific treatments or exposures. And for a scalar outcome $Y$, the ATE is defined as:
\begin{equation*}
	\Delta = E\left (Y^1\right) - E\left (Y^0\right),
\end{equation*}
where $Y^1$ and $Y^0$ are the potential outcomes under treatment or control, respectively~\citep{rubin1974estimating}. Under the Stable Unit Treatment Value Assumption~\citep{rubin1980randomization}, the observed outcome $Y$ is $Y=AY^1+(1-A)Y^0$. 

This framework can be extended to time-to-event outcomes. In this study, which involves competing risks, our outcome of interest is not a single scalar $Y$, but the CIF.
The cause-specific CIF for event $k$ for is defined as $F_{k}(t) = P(T <t , \epsilon = k)$.
Specifically, let $F_{k}^1(t)$ and $F_{k}^0(t)$ be the counterfactual CIF for event $k$ if individuals were assigned to the treatment ($A=1$) and the control ($A=0$), respectively. The causal effect of interest is the cause-specific CIF difference function~\citep{young2020causal}:
\begin{equation}\label{cif_diff}
	\Delta_k(t) = E\left[F_{k}^1(t)\right]-E\left[F_{k}^0(t)\right],\; t>0.
\end{equation}
This function $\Delta_k(t)$ represents the average causal effect of the treatment on the cause-specific CIF for event $k$ at any given time $t$.

\subsection{Pseudo-value for survival outcomes}
Standard causal methods, which we use to estimate our target estimand \eqref{cif_diff}, require a complete, scalar outcome for each individual. However, our time-to-event data is subject to right-censoring, which means the individual contribution to the CIF is not fully observed for all participants. Therefore, we first compute jackknife pseudo-values~\citep{andersen2003generalised}, where each pseudo-value provides an uncensored representation of the individual's contribution to the cause-specific CIF. Klein and Andersen~\cite{klein2005regression} extended pseudo-value approach into competing risk data, and proposed the pseudo-value of cause-specific CIF, which serves as a complete-data proxy for the $i$th individual's contribution to the of the cause-specific CIF.

We consider a sample of $n$ independent and identically distributed individuals, indexed by the subscript $i = 1, \ldots, n$. For each individual $i$, let $D_i=\{\tilde{T}_i,\tilde{\epsilon}_i,\boldsymbol{X}_i,A_i\}$ denote the observed data, and let $\boldsymbol{D} = (D_1, \ldots, D_n)$ represent the full observed dataset.
$\boldsymbol{D}^{(-i)}=(D_1,\dots,D_{i-1},D_{i+1},\dots,D_n)$ denote the ``leave-one-out'' sample excluding individual $i$. Then, the $i$th individual's pseudo-value of cause-specific CIF for event $k$ is given by
\begin{equation}\label{pseudo_CIF}
	\begin{split}
		F_k^{(i)}(t)=n \hat{F}_k(t)-(n-1)\hat{F}_k^{(-i)}(t),
	\end{split}
\end{equation}
where $\hat{F}_k(t)$ and $\hat{F}_k^{(-i)}(t)$ is the estimate of cause-specific CIF for event $k$ based on $\boldsymbol{D}$ and $\boldsymbol{D}^{(-i)}$, respectively.

Graw et al.~\cite{graw_pseudo-values_2009} proposed an approach to estimate the pseudo-value of cause-specific CIF for event $k$ with Aalen-Johansen estimator. They also demonstrated that $F_k^{(i)}(t)$ satisfy a conditional unbiasedness condition, and proved the consistency and asymptotic normality of regression coefficients estimated using this approach. The specific details regarding the estimation process are provided in Appendix \ref{app:pseudo_CIF_est}.

We therefore adopt $F_k^{(i)}(t)$ estimated by \eqref{pseudo_CIF} as the complete-data outcome variable in \eqref{cif_diff} and the naive estimator of $\Delta_k(t)$:
\begin{equation}
	\begin{split}
		\hat{\Delta}_k^{\text{naive}}(t)=\frac{1}{n}\sum_{i=1}^n\left[A_iF^{(i)}_k(t)-(1-A_i)F_k^{(i)}(t)\right].
	\end{split}
\end{equation}

\subsection{IPW adjusted pseudo-values}
In observational studies, directly estimating the difference of the cause-specific CIF with observed time of treatment and control groups may lead to a biased causal effect estimation due to unbalanced covariates (confounding) between the two groups. To deal with this, Rosenbaum and Rubin~\cite{rosenbaum1983central} proposed the PS method. They defined the propensity score $\pi(\boldsymbol{X})$ as the conditional probability of an individual $i$ receiving the treatment ($A=1$) given their observed covariates
\begin{equation*}
	\pi(\boldsymbol{X_i})=P(A_i=1|\boldsymbol{X_i}),
\end{equation*}
and the estimated value $\hat{\pi}(\boldsymbol{X})$ can be obtained by the logistic regression model. With strong ignorability ($\left(Y^1,Y^0\right) \perp A|\boldsymbol{X}$) and positivity ($0<\pi(\mathbf{X})<1$) assumption, they proved that $\pi(\boldsymbol{X})$ is sufficient to control for confounding. So, conditional on the propensity score, the distribution of the covariates $\boldsymbol{X}$ is theoretically identical between the treatment and control groups.

Then the IPW estimator of $\Delta_k(t)$ is given as~\citep{zeng2023propensity}:
\begin{equation}\label{IPW_CIF}
	\begin{split}
		\hat{\Delta}_{k}^{\text{IPW}}(t)=\frac{1}{n}\sum_{i=1}^n\left[\frac{A_i F_k^{(i)}(t)}{\hat{\pi}_i(\boldsymbol{X})}-\frac{(1-A_i) F_k^{(i)}(t)}{1-\hat{\pi}_i(\boldsymbol{X})}\right].
	\end{split}
\end{equation}

\subsection{Pseudo-value-based outcome regression}
G-computation via outcome regression modeling is a traditional method for estimating causal effects from observational studies. It directly models the outcome variable $Y$ as a function of both the treatment variable $A$ and a vector of observed confounders $\boldsymbol{X}$, which is the reason that it often be referred to as direct confounding adjustment. And this model has high statistical efficiency when the outcome model is correctly specified.

We can estimate the pseudo-value of cause-specific CIF for $i$th individual in \eqref{pseudo_CIF} at a set of distinct times $\boldsymbol{t}=\{t_1,\cdots,t_h\}$ by
\begin{align*}
	\hat{F}_k^{(i)}(t_j)=&n\hat{F}_k(t_j)-(n-1)\hat{F}_k^{(-i)}(t_j).
\end{align*} 
Using these pseudo-values as (pseudo) responses, we can use generalized estimating equations (GEE) to model the effects of covariates on outcome~\citep{klein2005regression}. Let $g(\cdot)$ be the link function (for example, $g(x)=\log\{-\log(1-x)\}$), the transformation model is 
\begin{equation*}
	g(F_k^{(i)}(t_j))=\alpha_k^{t_j}+\gamma_k A_i+\boldsymbol{\beta_k^T X_i},
\end{equation*}
where $\alpha_k^{t_j}$ is the intercept term for time $t_j$, and $\gamma_k$ and $\boldsymbol{\beta}_k$ are regression parameters. Note that, although $\alpha_k^{t_j}$ are time-specific, $\gamma_k$ and $\boldsymbol{\beta}_k$ are shared between different time points $\boldsymbol{t}=\{t_1,\dots, t_h\}$. The estimation of unknown parameters $\gamma_k$, $\boldsymbol{\beta}_k$ and $\Lambda_k=\{\alpha_k^{t_1},\dots \alpha_k^{t_h}\}$ are the solutions of the following generalized estimating equation:
\begin{equation*}
	\begin{split}
		U(\boldsymbol{\varphi_k})=\sum_{i=1}^n \frac{\partial g^{-1}(\boldsymbol{t,\varphi}_k;A_i,\boldsymbol{X_i})}{\partial\boldsymbol{\varphi}_k}V_i^{-1} \left[\hat{F}_k^{(i)}(\boldsymbol{t})-g^{-1}(\boldsymbol{t,\varphi}_k;A_i,\boldsymbol{X_i})\right]=0,
	\end{split}
\end{equation*}
where $\boldsymbol{\varphi}_k=(\Lambda_k,\gamma_k,\boldsymbol{\beta}_k)^T$, $\hat{F}^{(i)}_k(\boldsymbol{t})=(\hat{F}_k^{(i)}(t_1), \dots, \hat{F}_k^{(i)}(t_h))^T$ and $V_i^{-1}$ is a working covariance matrix which may account for the correlation structure inherent to pseudo-values. The OR estimator for $\Delta_k(t)$ based on pseudo-values is
\begin{equation*}
	\hat{\Delta}_k^{\text{OR}}(t)=\frac{1}{n}\sum_{i=1}^n\left[g^{-1}(\hat{\alpha^t_k}+\hat{\gamma_k}+\boldsymbol{\hat{\beta}^T_kX_i})-g^{-1}(\hat{\alpha^t_k}+\boldsymbol{\hat{\beta}^T_kX_i})\right],
\end{equation*}
where $g^{-1}(\cdot)$ is the inverse function of $g(\cdot)$.

\subsection{Multiply-robust estimator}
As previously discussed, the validity of doubly robust estimators is compromised under dual model misspecification. To address this critical gap, we propose a Multiply robust estimator to estimate the cause-specific CIF difference. Our estimator extends the foundational multiply robust framework of Han and Wang~\cite{han2013estimation} and adapts it specifically for the competing risks setting, utilizing the pseudo-value outcomes derived in \eqref{pseudo_CIF}.

To achieve multiple robustness, we specify multiple PS and OR models. In our proposed framework, candidate models refer to a user-specified collection of postulated working models for the propensity score and outcome regression, collectively utilized to safeguard the causal estimation against single-model misspecification. Let $\boldsymbol{Z}=(\boldsymbol{X},A)$ denote the combined covariates and treatment. We then suppose there are $L$ candidate models for the propensity score set $\mathcal{P}=\{p^l(\boldsymbol{X}),\quad l=1,2,3,\dots, L\}$, and $M$ candidate models for the outcome regression models set $\mathcal{Q}=\{q^m(\boldsymbol{Z}), \quad m=1,2,3,\dots, M\}$, where $p^l(\boldsymbol{X})=\pi^l(\boldsymbol{X})$, $q^m(\boldsymbol{Z_i})=g(F_k^{(i)}(t)|\boldsymbol{Z_i})$. The total number of candidate models is $S=L+M$. Let $\mathcal{I}=\{i : A_i=1\}$ denote the set of indices for treated subjects and $\mathcal{J}=\{i : A_i=0\}$ denote the set for control subjects, with respective sample sizes $n_1=|\mathcal{I}|$ and $n_0=|\mathcal{J}|$. Our proposed multiply robust estimator for the cause-specific CIF difference, $\Delta_k(t)$, is then defined as a weighted estimator:
\begin{equation*}
	\hat{\Delta}_k^{\text{MR}}(t)=\sum_{i\in \mathcal{I}}\hat{w}_iF_k^{(i)}(t)-\sum_{j\in \mathcal{J}}\hat{w}_j\hat{F}_k^{(j)}(t),
\end{equation*}
where $\hat{w}_i$ and $\hat{w}_j$ are subject-specific weights. These weights are constructed by integrating the information from the entire collection of $S$ models in $\mathcal{P}$ and $\mathcal{Q}$.

Theoretically, proposed estimator achieves multiple robustness through a specialized weight construction mechanism based on empirical likelihood. Rather than relying on estimated propensity scores or predicted responses from a single regression model, we treat the subject-specific weights as unknown parameters and estimate them by maximizing the empirical likelihood subject to a system of multiple calibration constraints. Crucially, these constraints compel the reweighted empirical distribution to simultaneously satisfy the moment conditions implied by all $S = L + M$ candidate models in $\mathcal{P}$ and $\mathcal{Q}$. By calibrating against the entire candidate set, the estimator successfully integrates information across all postulated models. Consequently, the estimator is guaranteed to remain consistent and asymptotically unbiased for the true cause-specific CIF difference (1), provided that any single model among the $L$ PS models or $M$ OR models is correctly specified. Furthermore, it has been shown that if at least one $p^l(X) \in \mathcal{P}$ and at least one $q^m(Z) \in \mathcal{Q}$ are correctly specified, the estimator attains the semiparametric efficiency bound, circumventing the need for a priori knowledge of which specific models are correct~\cite{han2013estimation}.

Using Han's method~\citep{han2014further}, we can estimate $w_i$ by maximizing $\prod_{i\in \mathcal{I}}w_i$ subject to the following constraints:
\begin{align}\label{constriants_i}
	&\ 0 \le w_i \le 1,\ \text{for all }i\in \mathcal{I}, \nonumber \\ 
	&\sum_{i\in \mathcal{I}}w_i=1, \nonumber \\
	&\sum_{i\in \mathcal{I}}w_i\left[\hat{p}^l(\boldsymbol{X}_i)-\hat{\theta}_1^l\right]=0,\ l=1,2,3,\dots,L, \nonumber \\
	&\sum_{i\in \mathcal{I}}w_i\left[\hat{q}_1^m(\boldsymbol{Z}_i)-\hat{\eta}_1^m\right]=0,\ m=1,2,3,\dots,M,
\end{align}
where $\hat{\theta}_1^l=n_1^{-1}\sum_{i\in \mathcal{I}}p^l(\boldsymbol{X}_i)$ and $\hat{\eta}_1^m=n_i^{-1}\sum_{i\in \mathcal{I}}^n q_1^m(\boldsymbol{Z}_i)$. 
Similarly, we can also estimate $w_j$ by maximizing $\prod_{j\in\mathcal{J}}w_j$ subject to the following constraints:
\begin{align}\label{constriants_i}
	&\ 0 \le w_j\le 1,\ \text{for all }j\in\mathcal{J}, \nonumber \\
	&\sum_{j\in\mathcal{J}}w_j=1, \nonumber \\
	&\sum_{j\in\mathcal{J}}w_j\left\{\left[1-\hat{p}^l(\boldsymbol{X}_j)\right]-\hat{\theta}_0^l\right\}=0,\ l=1,2,3,\dots,L, \nonumber \\
	&\sum_{j\in \mathcal{J}}w_j\left[\hat{q}_0^m(\boldsymbol{Z}_j)-\hat{\eta}_0^m\right]=0,\ m=1,2,3,\dots,M,
\end{align}
where $\hat{\theta}_0^l=n_0^{-1}\sum_{j\in \mathcal{J}}^n\left[1-p^l(\boldsymbol{X}_j)\right]$ and $\hat{\eta}_0^m=n_0^{-1}\sum_{j\in \mathcal{J}}^nq_0^m(\boldsymbol{Z}_j)$. 

Write 
\begin{align*}
	\hat{g}_1(\boldsymbol{Z})^T=\{\hat{\pi}^1(\boldsymbol{X})-\hat{\theta}_1^1,\dots,\hat{\pi}^L(\boldsymbol{X})-\hat{\theta}_1^L, \hat{q}_1^1(\boldsymbol{Z})-\hat{\eta}_1^1,\dots,\hat{q}_1^M(\boldsymbol{Z})-\hat{\eta}_1^M\},\\
	\hat{g}_0(\boldsymbol{Z})^T=\{\hat{\pi}^1(\boldsymbol{X})-\hat{\theta}_0^1,\dots,\hat{\pi}^L(\boldsymbol{X})-\hat{\theta}_0^L, \hat{q}_0^1(\boldsymbol{Z})-\hat{\eta}_0^1,\dots,\hat{q}_0^M(\boldsymbol{Z})-\hat{\eta}_0^M\}.
\end{align*}
Using the Lagrange multiplier method, which is routinely employed in the empirical likelihood theory, we can obtain 
\begin{align*}
	\hat{w}_i &= \frac{1}{n_1} \frac{1}{1 + \hat{\rho}_1^T \hat{\boldsymbol{g}}_1(\mathbf{Z}_i)} \Bigg/ \left\{ \frac{1}{n_1} \sum_{i\in\mathcal{I}} \frac{1}{1 + \hat{\rho}_1^T \hat{\boldsymbol{g}}_1(\mathbf{Z}_i)} \right\} \quad \text{for all }i\in \mathcal{I},\\
	\hat{w}_j &= \frac{1}{n_0} \frac{1}{1 + \hat{\rho}_0^T \hat{\boldsymbol{g}}_0(\mathbf{Z}_j)} \Bigg/ \left\{ \frac{1}{n_0} \sum_{j \in \mathcal{J}} \frac{1}{1 + \hat{\rho}_0^T \hat{\boldsymbol{g}}_0(\mathbf{Z}_j)} \right\} \quad \text{for all }j\in \mathcal{J},
\end{align*}
and $\hat{\rho}_0^T=(\hat{\rho}_{01},\hat{\rho}_{02},\dots,\hat{\rho}_{0S})$, $\hat{\rho}_1^T=(\hat{\rho}_{11},\hat{\rho}_{12},\dots,\hat{\rho}_{1S})$ can be solved by 
\begin{align*}
	\frac{1}{n_1}\sum_{i\in\mathcal{I}}\frac{\hat{\boldsymbol{g}}_1(\boldsymbol{Z}_i)}{1+\hat{\boldsymbol{\rho}}_1^T\hat{\boldsymbol{g}}_1(\boldsymbol{Z}_i)}&=0,\\
	\frac{1}{n_0}\sum_{j\in\mathcal{J}}\frac{\hat{\boldsymbol{g}}_0(\boldsymbol{Z}_j)}{1+\hat{\boldsymbol{\rho}}_0^T\hat{\boldsymbol{g}}_0(\boldsymbol{Z}_j)}&=0.
\end{align*}

Due to the non-negativity of $w_i$ and $w_j$, $\hat{\rho}_1^T$ and $\hat{\rho}_0^T$ must satisfy inequalities $1+\hat{\boldsymbol{\rho}}_1^T\hat{\boldsymbol{g}}_1(\boldsymbol{Z}_i)>0$ and $1+\hat{\boldsymbol{\rho}}_0^T\hat{\boldsymbol{g}}_0(\boldsymbol{Z}_j)>0$. The estimation of $w_i$ and $w_j$ can be solved using the Newton-Raphson algorithm.

\subsection{Bootstrap}
Han~\cite{han2016combining} proved that the estimator follows an asymptotic normal distribution and provided an expression for its asymptotic variance, under the assumption that the propensity score model is correctly specified. Given the practical difficulties in applying this approach, the bootstrap method is recommended to estimate confidence interval.
Therefore, we use three types of bootstrap confidence interval methods to estimate confidence interval (CI).

Specifically, $n$ subjects are resampled from the original data with the replacement for $B$ times to obtain $B$ bootstrap samples. Let $\hat{\Delta}^b(t)$ be the estimated difference of cause-specific CIF from the $b$-th bootstrap sample, $b=1,\dots,B$. Then the bootstrap variance estimator for $\hat{\Delta}^b(t)$ is defined as
\begin{equation*}
	\hat{\mathrm{Var}}\left[\hat{\Delta}^b(t)\right]=\frac{1}{B-1}\sum_{b=1}^B\left[\hat{\Delta}^b(t)-\frac{1}{B}\sum_{b=1}^B\hat{\Delta}^b(t)\right]^2.
\end{equation*}
Under the significance level $\alpha$, a normality-based confidence interval for $\hat{\Delta}(t)$ is 
\begin{equation*}
	\mathrm{CI_{nor}} = \hat{\Delta}^b(t) \pm z_{\alpha/2}\cdot \sqrt{\hat{\mathrm{Var}}\left[\hat{\Delta}^b(t)\right]},
\end{equation*}
where $z_{\alpha/2}$ is the upper $\alpha/2$ quantile of the standard normal distribution.

Besides, the percentile-based $(1-\alpha)\cdot100\%$ confidence interval is 
\begin{equation*}
	\mathrm{CI_{per}} = \left[\hat{\Delta}^*(t)_{(\alpha/2)},\ \hat{\Delta}^*(t)_{(1-\alpha/2)}\right],
\end{equation*}
where $\hat{\Delta}^*(t)_q$ is the $q$-th quantile of the Bootstrap empirical distribution. A $(1-\alpha)\cdot100\%$ pivotal confidence interval is
\begin{equation*}
	\mathrm{CI_{piv}} = \left[2\hat{\Delta}(t)-\hat{\Delta}^*(t)_{(1-\alpha/2)},\ 2\hat{\Delta}(t)-\hat{\Delta}^*(t)_{(\alpha/2)}\right]. 
\end{equation*}

\section{Results}
\subsection{Simulation Analysis}
In this section, we conduct extensive simulation studies under various scenarios to evaluate the performance of estimators for the CIF difference. We compare our proposed multiply robust estimator with traditional estimators relying solely on either a PS model or an OR model, as well as standard doubly robust estimators.
\subsubsection{Simulation Set-Up}
We generated a 15-dimensional covariate vector $\boldsymbol{X}=(\boldsymbol{X^{(1)}, \boldsymbol{X^{(2)}}, \boldsymbol{X^{(3)}}})$ for each subject. The components of $\boldsymbol{X}$ were generated independently from the following distributions:
\begin{align*}
	\boldsymbol{X^{(1)}}&=(X_1, X_2, \dots, X_5) \sim \mathcal{N}_5(\boldsymbol{0},\boldsymbol{\Sigma}), \quad \text{where } \Sigma_{ij} = 0.5^{|i-j|},\\
	\boldsymbol{X^{(2)}}&=(X_6, X_7, \dots, X_{14}) \sim \mathcal{N}_9(\boldsymbol{0},\boldsymbol{I_9}),\\
	\boldsymbol{X^{(3)}}&=X_{15} \sim \text{Bernoulli(0.5)}.
\end{align*}
Here, $\boldsymbol{\Sigma}$ represents the covariance matrix with an AR(1) structure, and $\boldsymbol{I_9}$ denotes a $9\times9$ identity matrix. 
A binary exposure $A$ was simulated from a Bernoulli distribution according to the following propensity score 

$$\text{logit}\left[\pi(\boldsymbol{X})\right]=0.5X_1^2+0.3X_2-0.6X_6+0.5X_{15}.$$

Assuming the number of competing events to be $K=2$, we simulated the survival times from a Gompertz distribution and specified the proportional hazards model for the cause-specific hazard of each event $k$ as:
\begin{equation*}           h_k(t)=a_k\exp\left(\text{LP}_k+\lambda_k A\right)\exp(\rho_kt),\;k=1,2\;,
\end{equation*}
where $a_k$ is a constant baseline hazard, $\lambda_k$ is the event-specific treatment effect on the log hazard, $\text{LP}_k$ is the linear predictor of covariate effects on the log hazard, and $\rho_k$ is the time effect on the log hazard. We set $\lambda_k=1$, $a_k=0.5$, and $\rho_k=1.5$ for both event types, and defined the linear predictors
\begin{align*}
	\text{LP}_1&=0.4X_1^2+0.2X_2-0.7X_6+0.5X_{15},\\
	\text{LP}_2&=-0.3X_1^2-0.2X_2+0.6X_6-0.4X_{15},
\end{align*}
for the first and the second competing events, respectively. According to Bender et al.~\cite{bender2005generating}, the cause-specific survival time $T_k$ for each event was generated as:
\begin{equation*}
	T_{k} = \frac{1}{\rho_k}\log\left[1-\frac{\rho_k \log(U_k)}{a_k\exp(\text{LP}_k+\lambda_k A)}\right],\;k=1,2\;,
\end{equation*}
where $U_k$ was drawn from a standard uniform distribution. 

To test the robustness of the proposed estimator under various censoring proportions, we assumed the censoring time $C$ followed a uniform distribution $U(c_{min}, c_{max})$ and adjusted the parameters $c_{min}$ and $c_{max}$ to achieve the desired censoring proportions of 0\%, 10\%, and 25\%. The observed event time was $\tilde{T}=\min(T_1, T_2, C)$, and the observed status $\tilde{\epsilon}$ was defined as $\tilde{\epsilon}=0$ if $\tilde{T}=C$ (censored), $\tilde{\epsilon}=1$ if $\tilde{T}=T_1$, and $\tilde{\epsilon}=2$ if $\tilde{T}=T_2$.

To establish the ground truth for performance evaluation, the true causal effect was calculated using the cause-specific survival times $T_1$ and $T_2$ without censoring. The ultimately observed data for each simulated subject comprised the 15-dimensional covariate vector $\boldsymbol{X}$, the binary exposure $A$, the observed survival time $\tilde{T}$, and the observed event status indicator $\tilde{\epsilon}$. Based on these simulated datasets $\boldsymbol{O}=(\boldsymbol{X},A,\tilde{T},\tilde{\epsilon})$, we designed several scenarios to evaluate the finite-sample performance and the multiple robustness property of the proposed estimator.

\textbf{Scenario I:} In this scenario, the correctly specified models, denoted as $p^1$ and $q^1$, included the true functional forms and all relevant covariates:
\begin{equation*}
	\begin{cases}
		p^1:\pi^{1}(\mathbf{X}) \sim \left(X_1^2, X_2,\dots,X_{15}\right) , & \\
		q^1:q^{1}(\mathbf{X}, A) \sim \left(X_1^2, X_2, \dots, X_{15}, A\right).& 
	\end{cases}
\end{equation*}
To assess the multiple robustness property, we intentionally constructed misspecified models by restricting the candidate set to a subset of the 15 covariates, that is ($X_1, \dots, X_5,X_{15}$), effectively omitting some key variables responsible for generating the survival times. Furthermore, we misspecified the functional form of $X_1$ by replacing its true quadratic relationship ($X_1^2$) with a linear term. Thus, the misspecified PS model ($p^2$) and OR model ($q^2$) were given by:
\begin{equation*}
	\begin{cases}
		p^2:\pi^{2}(\mathbf{X}) \sim \left(X_1, \dots, X_5, X_{15}\right), & \\
		q^2:q^{2}(\mathbf{X}, A) \sim \left(X_1, \dots, X_5, X_{15}, A\right).& 
	\end{cases}
\end{equation*}
Finally, we evaluated the performance of the proposed estimator under various model combinations drawn from the set $\{p^1, p^2, q^1, q^2\}$.

\textbf{Scenario II:} In this scenario, we retained the correctly specified models ($p^1$ and $q^1$) defined in Scenario I. To introduce a new form of model misspecification, we once again intentionally restricted the covariate candidate set, but selected a different subset ($X_1,\dots,X_5$). Furthermore, we replaced the true quadratic relationship of $X_1$ ($X_1^2$) with a sinusoidal transformation, $\sin(X_1)$. Thus, these new misspecified models for the PS ($p^3$) and OR ($q^3$) were given by:
\begin{equation*}
	\begin{cases}
		p^3:\pi^{3}(\mathbf{X}) \sim \left(\sin(X_1),X_2,\dots,X_5\right), & \\
		q^3:q^{3}(\mathbf{X}, A) \sim \left(\sin(X_1),X_2,\dots,X_5, A\right).& 
	\end{cases}
\end{equation*}
Consequently, the expanded candidate set comprised the correctly specified models ($p^1, q^1$) alongside two distinct types of misspecified models ($p^2, q^2$ and $p^3, q^3$). We then evaluated the finite-sample performance of the proposed estimator under all possible model combinations drawn from the comprehensive set $\{p^1, p^2, p^3, q^1, q^2, q^3\}$.

\textbf{Scenario III:} Departing from the manual variable omission strategies used in Scenarios I and II, Scenario III utilized a data-driven approach by employing LASSO~\citep{tibshirani1996regression} for objective variable selection. In this setting, both the correctly specified and misspecified models were constructed by applying LASSO to the full suite of 15 candidate covariates. For the correctly specified PS ($p'^1$) and OR ($q'^1$) models, variables were selected from the sets $\{X_1^2, X_2, \dots, X_{15}\}$ and $\{A, X_1^2, X_2, \dots, X_{15}\}$, respectively. For the misspecified models ($p'^2$ and $q'^2$), LASSO screened variables from candidate sets where the true quadratic relationship ($X_1^2$) was replaced by a linear main effect ($X_1$), namely from $\{X_1, X_2, \dots, X_{15}\}$ and $\{A, X_1, X_2, \dots, X_{15}\}$.
We evaluated the performance of the proposed estimator under various model combinations drawn from the set $\{p'^1, p'^2, q'^1, q'^2\}$.

We used a sample size of $n=500$ and generated 500 replicate datasets for the simulation. For each dataset, we took 200 bootstrap iterations and calculated evaluation criteria including mean bias, mean square error (MSE), empirical standard deviation (SD), average estimated standard error ($\overline{\text{SE}}$) and confidence interval coverage rates. Bias and MSE evaluate the deviation of the estimates from the true parameter, while the coverage rate assesses the reliability of the confidence intervals. We constructed intervals using the normal (CR.nor), pivotal (CR.piv), and percentile (CR.per) methods.

\subsubsection{Simulation Results}
Our simulation results are shown in Table \ref{tab:result_model1} - \ref{tab:result_model_lasso} and Supplementary Tables S1-S6. In the simulation results, the $i$th single propensity score model or outcome regression model is represented by $p^i$ or $q^i$, respectively. Multiply robust estimators composed of two kinds of models are represented by ``MR$p^ip^jq^iq^j$'' and the name is decided by propensity score models and outcome regression models included in multiply robust estimators. For example, the first four estimators in Table \ref{tab:result_model1} ($p^1$, $p^2$, $q^1$, $q^2$) constitute the single-model specifications group. The fifth to eighth estimators (MR$p^1$, MR$p^2$, MR$q^1$, MR$q^2$) form the single-model MR specifications group, where each estimator is obtained under the MR framework using only one model. The subsequent six estimators (MR$p^1p^2$, MR$p^1q^1$, MR$p^1q^2$, MR$p^2q^1$, MR$p^2q^2$, MR$q^1q^2$) represent the two-model MR combinations, which correspond to the conventional doubly robust estimators. The final five estimators constitute the three and four multi-model MR combinations group.

\begin{table}[h]
\caption{Simulation results of selected models $\{p^1,p^2,q^1,q^2\}$ using data with 10\% and 25\% censoring rates\label{tab:result_model1}}
\setlength{\tabcolsep}{1pt} 
\footnotesize
\begin{tabular*}{\textwidth}{@{\extracolsep\fill}lcccccccccc}
    \hline
    & \multicolumn{5}{@{}c@{}}{10\% censoring rate} & \multicolumn{5}{@{}c@{}}{25\% censoring rate} \\\cmidrule(lr){2-6}\cmidrule(lr){7-11}%
    Estimators & Bias & MSE & SD & $\overline{\text{SE}}$ & CR & Bias & MSE & SD & $\overline{\text{SE}}$ & CR \\
    \hline
    $p^1$ & $-$0.0578 & 0.7006 & 0.0836 & 0.0861 & 95.6 & $-$0.1968 & 0.7636 & 0.0852 & 0.0918 & 97.4\\
    $p^2$ & \textbf{1.4729} & \textbf{2.3583} & \textbf{0.0435} & \textbf{0.0448} & \textbf{11.2} & \textbf{1.2918} & \textbf{1.8648} & \textbf{0.0443} & \textbf{0.0492} & \textbf{23.8}\\
    $q^1$ & $-$0.0446 & 0.2290 & 0.0477 & 0.0563 & 97.2 & $-$0.1953 & 0.2683 & 0.0480 & 0.0656 & 97.2\\
    $q^2$ & \textbf{1.4517} & \textbf{2.2987} & \textbf{0.0438} & \textbf{0.0457} & \textbf{10.8} & \textbf{1.2672} & \textbf{1.8043} & \textbf{0.0446} & \textbf{0.0501} & \textbf{26.4}\\
    MR$p^1$ & $-$0.0423 & 0.2707 & 0.0519 & 0.0524 & 94.2 & $-$0.1851 & 0.3189 & 0.0534 & 0.0580 & 94.4\\
    MR$p^2$ & \textbf{1.4594} & \textbf{2.3194} & \textbf{0.0436} & \textbf{0.0447} & \textbf{11.2} & \textbf{1.2769} & \textbf{1.8278} & \textbf{0.0445} & \textbf{0.0491} & \textbf{24.2}\\
    MR$q^1$ & $-$0.0932 & 0.2482 & 0.0490 & 0.0510 & 96.6 & $-$0.2588 & 0.3151 & 0.0499 & 0.0570 & 95.2\\
    MR$q^2$ & \textbf{1.4519} & \textbf{2.3018} & \textbf{0.0441} & \textbf{0.0451} & \textbf{11.8} & \textbf{1.2654} & \textbf{1.8027} & \textbf{0.0449} & \textbf{0.0494} & \textbf{25.2}\\
    MR$p^1p^2$ & $-$0.0248 & 0.2677 & 0.0517 & 0.0525 & 94.2 & $-$0.1689 & 0.3097 & 0.0531 & 0.0581 & 94.6\\
    MR$p^1q^1$ & $-$0.0778 & 0.2615 & 0.0506 & 0.0512 & 95.4 & $-$0.2170 & 0.3182 & 0.0521 & 0.0568 & 94.6\\
    MR$p^1q^2$ & $-$0.0298 & 0.2677 & 0.0517 & 0.0525 & 94.4 & $-$0.1727 & 0.3122 & 0.0532 & 0.0580 & 94.4\\
    MR$p^2q^1$ & $-$0.0946 & 0.2485 & 0.0490 & 0.0523 & 95.8 & $-$0.1727 & 0.3122 & 0.0532 & 0.0580 & 94.4\\
    MR$p^2q^2$ & \textbf{1.4533} & \textbf{2.3036} & \textbf{0.0438} & \textbf{0.0448} & \textbf{11.6} & \textbf{1.2718} & \textbf{1.8172} & \textbf{0.0447} & \textbf{0.0493} & \textbf{24.6}\\
    MR$q^1q^2$ & $-$0.0897 & 0.2519 & 0.0494 & 0.0514 & 96.2 & $-$0.2538 & 0.3148 & 0.0501 & 0.0573 & 94.6\\
    MR$p^1p^2q^1$ & $-$0.0762 & 0.2561 & 0.0501 & 0.0513 & 95.2 & $-$0.2127 & 0.3104 & 0.0515 & 0.0568 & 94.6\\
    MR$p^1p^2q^2$ & $-$0.0063 & 0.2667 & 0.0517 & 0.0525 & 94.2 & $-$0.1494 & 0.3041 & 0.0531 & 0.0580 & 94.4\\
    MR$p^1q^1q^2$ & $-$0.0740 & 0.2605 & 0.0505 & 0.0513 & 95.2 & $-$0.2114 & 0.3148 & 0.0520 & 0.0569 & 94.6\\
    MR$p^2q^1q^2$ & $-$0.0927 & 0.2622 & 0.0504 & 0.0527 & 96.0 & $-$0.2456 & 0.3321 & 0.0522 & 0.0586 & 95.4\\
    MR$p^1p^2q^1q^2$ & $-$0.0709 & 0.2599 & 0.0505 & 0.0515 & 95.2 & $-$0.2061 & 0.3139 & 0.0522 & 0.0570 & 94.8\\
    \hline
\end{tabular*}
\vspace{1ex}
{\footnotesize
	\noindent\textsuperscript{a} Bias: mean bias ($\times 10$); MSE: mean squared error ($\times 10^2$); SD: standard deviation; $\overline{\text{SE}}$: average estimated standard error; CR: 95\% normality-based confidence interval coverage rate (\%).\\[0.5ex]
	\textsuperscript{b} Values in bold indicate results where all component working models are misspecified.
}
\end{table}
\begin{table}[h]
\caption{Partial simulation results of selected models $\{p^1,p^2,p^3,q^1,q^2, q^3\}$ using data with 10\% and 25\% censoring rates\label{tab:result_model2}}
\setlength{\tabcolsep}{2pt} 
\footnotesize
\begin{tabular*}{\textwidth}{@{\extracolsep\fill}lcccccccccc}
    \hline
    & \multicolumn{5}{@{}c@{}}{10\% censoring rate} & \multicolumn{5}{@{}c@{}}{25\% censoring rate} \\\cmidrule(lr){2-6}\cmidrule(lr){7-11}%
    Estimators & Bias & MSE & SD & $\overline{\text{SE}}$ & CR & Bias & MSE & SD & $\overline{\text{SE}}$ & CR \\
    \hline
    MR$p^2p^3$ & \textbf{1.4605} & \textbf{2.3237} & \textbf{0.0437} & \textbf{0.0448} & \textbf{11.2} & \textbf{1.2778} & \textbf{1.8309} & \textbf{0.0446} & \textbf{0.0492} & \textbf{24.0}\\
    MR$p^2q^2$ & \textbf{1.4533} & \textbf{2.3036} & \textbf{0.0438} & \textbf{0.0448} & \textbf{11.6} & \textbf{1.2718} & \textbf{1.8172} & \textbf{0.0447} & \textbf{0.0493} & \textbf{24.6}\\
    MR$p^2q^3$ & \textbf{1.4557} & \textbf{2.3105} & \textbf{0.0438} & \textbf{0.0448} & \textbf{11.6} & \textbf{1.2731} & \textbf{1.8207} & \textbf{0.0448} & \textbf{0.0492} & \textbf{23.8}\\
    MR$p^3q^2$ & \textbf{1.4528} & \textbf{2.3060} & \textbf{0.0442} & \textbf{0.0451} & \textbf{11.2} & \textbf{1.2680} & \textbf{1.8115} & \textbf{0.0452} & \textbf{0.0495} & \textbf{25.2}\\
    MR$p^3q^3$ & \textbf{1.5618} & \textbf{2.6308} & \textbf{0.0438} & \textbf{0.0441} & \textbf{7.6} & \textbf{1.3764} & \textbf{2.0944} & \textbf{0.0448} & \textbf{0.0485} & \textbf{17.6}\\
    MR$q^2q^3$ & \textbf{1.4502} & \textbf{2.2963} & \textbf{0.0440} & \textbf{0.0451} & \textbf{11.4} & \textbf{1.2650} & \textbf{1.8005} & \textbf{0.0448} & \textbf{0.0495} & \textbf{24.8}\\
    MR$p^2p^3q^2$ & \textbf{1.4510} & \textbf{2.2990} & \textbf{0.0440} & \textbf{0.0449} & \textbf{11.6} & \textbf{1.2695} & \textbf{1.8128} & \textbf{0.0449} & \textbf{0.0493} & \textbf{25.0}\\
    MR$p^2p^3q^3$ & \textbf{1.4563} & \textbf{2.3137} & \textbf{0.0440} & \textbf{0.0448} & \textbf{11.4} & \textbf{1.2740} & \textbf{1.8238} & \textbf{0.0448} & \textbf{0.0493} & \textbf{23.6}\\
    MR$p^2q^2q^3$ & \textbf{1.4526} & \textbf{2.3020} & \textbf{0.0439} & \textbf{0.0449} & \textbf{11.8} & \textbf{1.2710} & \textbf{1.8158} & \textbf{0.0448} & \textbf{0.0493} & \textbf{24.2}\\
    MR$p^3q^2q^3$ & \textbf{1.4522} & \textbf{2.3024} & \textbf{0.0440} & \textbf{0.0451} & \textbf{11.4} & \textbf{1.2701} & \textbf{1.8144} & \textbf{0.0449} & \textbf{0.0495} & \textbf{24.2}\\
    MR$p^1p^2p^3q^2$ & $-$0.0015 & 0.2679 & 0.0518 & 0.0527 & 93.8 & $-$0.1449 & 0.3052 & 0.0534 & 0.0581 & 94.4\\
    MR$p^1p^2p^3q^3$ & 0.0006 & 0.2649 & 0.0515 & 0.0526 & 94.0 & $-$0.1428 & 0.3017 & 0.0531 & 0.0580 & 94.0\\
    MR$p^1p^2q^2q^3$ & $-$0.0005 & 0.2665 & 0.0517 & 0.0526 & 93.8 & $-$0.1445 & 0.3030 & 0.0532 & 0.0580 & 94.8\\
    MR$p^1p^3q^2q^3$ & 0.0007 & 0.2650 & 0.0515 & 0.0525 & 94.2 & $-$0.1431 & 0.3014 & 0.0531 & 0.0580 & 94.6\\
    MR$p^2p^3q^1q^2$ & $-$0.0905 & 0.2614 & 0.0504 & 0.0527 & 95.8 & $-$0.2454 & 0.3295 & 0.0519 & 0.0586 & 95.2\\
    MR$p^2p^3q^1q^3$ & $-$0.0936 & 0.2588 & 0.0501 & 0.0526 & 96.2 & $-$0.2462 & 0.3269 & 0.0516 & 0.0586 & 94.8\\
    MR$p^2p^3q^2q^3$ & \textbf{1.4481} & \textbf{2.2919} & \textbf{0.0442} & \textbf{0.0453} & \textbf{12.2} & \textbf{1.2654} & \textbf{1.8047} & \textbf{0.0451} & \textbf{0.0497} & \textbf{25.4}\\
    MR$p^2q^1q^2q^3$ & $-$0.0905 & 0.2593 & 0.0502 & 0.0526 & 96.0 & $-$0.2415 & 0.3286 & 0.0520 & 0.0585 & 95.0\\
    MR$p^3q^1q^2q^3$ & $-$0.0932 & 0.2562 & 0.0498 & 0.0527 & 96.0 & $-$0.2441 & 0.3253 & 0.0516 & 0.0586 & 95.4\\
    MR$p^1p^2p^3q^1q^2$ & $-$0.0716 & 0.2603 & 0.0506 & 0.0517 & 95.2 & $-$0.2069 & 0.3136 & 0.0521 & 0.0571 & 94.8\\
    MR$p^1p^2p^3q^1q^3$ & $-$0.0734 & 0.2538 & 0.0499 & 0.0516 & 95.2 & $-$0.2075 & 0.3066 & 0.0514 & 0.0570 & 94.6\\
    MR$p^1p^2p^3q^2q^3$ & 0.0065 & 0.2683 & 0.0518 & 0.0534 & 95.0 & $-$0.1416 & 0.3089 & 0.0538 & 0.0588 & 94.8\\
    MR$p^1p^2q^1q^2q^3$ & $-$0.0712 & 0.2568 & 0.0502 & 0.0516 & 95.2 & $-$0.2060 & 0.3109 & 0.0519 & 0.0570 & 95.4\\
    MR$p^1p^3q^1q^2q^3$ & $-$0.0726 & 0.2543 & 0.0500 & 0.0516 & 95.6 & $-$0.2058 & 0.3071 & 0.0515 & 0.0570 & 95.4\\
    MR$p^2p^3q^1q^2q^3$ & $-$0.0896 & 0.2553 & 0.0498 & 0.0529 & 95.6 & $-$0.2413 & 0.3267 & 0.0519 & 0.0589 & 95.4\\
    MR$p^1p^2p^3q^1q^2q^3$ & $-$0.0670 & 0.2558 & 0.0502 & 0.0523 & 95.6 & $-$0.2033 & 0.3117 & 0.0520 & 0.0578 & 95.2\\
    \hline
\end{tabular*}
\vspace{1ex}
{\footnotesize
	\noindent\textsuperscript{a} Bias: mean bias ($\times 10$); MSE: mean squared error ($\times 10^2$); SD: standard deviation; $\overline{\text{SE}}$: average estimated standard error; CR: 95\% normality-based confidence interval coverage rate (\%).\\[0.5ex]
	\textsuperscript{b} Values in bold indicate results where all component working models are misspecified.
}
\end{table}

As shown in Table \ref{tab:result_model1}, results marked in bold correspond to estimators relying exclusively on misspecified models. This category includes estimates based on a single incorrect model (e.g., $p^2$, $q^2$) as well as fully misspecified multiply robust estimators (e.g., MR$p^2q^2$). Consistent with expectations, these estimators yielded significantly biased estimates with high MSE, and their 95\% CI coverage rates fell significantly below the nominal 95\% level. In contrast, other multiply robust estimators (e.g., MR$p^1p^2q^1$, MR$p^1p^2q^2$, MR$p^1q^1q^2$, MR$p^2q^1q^2$, MR$p^1p^2q^1q^2$) consistently produced MSEs near zero and coverage rates close to the 0.95 level, even when multiple misspecified PS and OR models were included in the candidate set. These findings indicate that the multiply robust estimators perform well provided that at least one component model (either PS or OR) is correctly specified. 
Simulation results for combinations involving six candidate models, as detailed in Table \ref{tab:result_model2} and Supplementary Tables S1-S3, yield patterns consistent with those observed in Tabel \ref{tab:result_model1}. 

Furthermore, similar patterns emerged in Table \ref{tab:result_model_lasso} (Scenario III), which incorporated LASSO for data-driven variable selection prior to model estimation. Despite the added complexity of the selection process, the multiply robust estimators continued to exhibit minimal bias and appropriate 95\% CI coverage rates. These results underscore the flexibility of our proposed framework, demonstrating that it seamlessly accommodates various model-building and variable selection algorithms. Importantly, the application of different selection methods does not compromise the underlying multiple robustness of the estimators, ensuring reliable inference in practical settings.


\begin{table}[h]
\caption{Simulation results of models $\{p'^1,p'^2,q'^1,q'^2\}$ selected by LASSO using data with 10\% and 25\% censoring rates\label{tab:result_model_lasso}}
\setlength{\tabcolsep}{1pt} 
\footnotesize
\begin{tabular*}{\textwidth}{@{\extracolsep\fill}lcccccccccc}
    \hline
    & \multicolumn{5}{@{}c@{}}{10\% censoring rate} & \multicolumn{5}{@{}c@{}}{25\% censoring rate} \\\cmidrule(lr){2-6}\cmidrule(lr){7-11}%
    Estimators & Bias & MSE & SD & $\overline{\text{SE}}$ & CR & Bias & MSE & SD & $\overline{\text{SE}}$ & CR \\
    \hline
    $p'^1$ & $-$0.0536 & 0.6510 & 0.0806 & 0.0855 & 96.2 & $-$0.1923 & 0.7149 & 0.0824 & 0.0913 & 97.2\\
    $p'^2$ & \textbf{0.6328} & \textbf{0.6236} & \textbf{0.0473} & \textbf{0.0533} & \textbf{78.2} & \textbf{0.4706} & \textbf{0.4633} & \textbf{0.0492} & \textbf{0.0584} & \textbf{88.8}\\
    $q'^1$ & $-$0.0558 & 0.2220 & 0.0468 & 0.0530 & 96.8 & $-$0.2039 & 0.2666 & 0.0475 & 0.0598 & 96.6\\
    $q'^2$ & \textbf{0.5878} & \textbf{0.5511} & \textbf{0.0454} & \textbf{0.0510} & \textbf{80.4} & \textbf{0.4230} & \textbf{0.3950} & \textbf{0.0465} & \textbf{0.0570} & \textbf{92.8}\\
    MR$p'^1$ & $-$0.0416 & 0.2651 & 0.0514 & 0.0524 & 95.0 & $-$0.1840 & 0.3148 & 0.0531 & 0.0580 & 94.4\\
    MR$p'^2$ & \textbf{0.6118} & \textbf{0.5781} & \textbf{0.0452} & \textbf{0.0489} & \textbf{78.8} & \textbf{0.4477} & \textbf{0.4188} & \textbf{0.0468} & \textbf{0.0541} & \textbf{90.2}\\
    MR$q'^1$ & $-$0.0905 & 0.2564 & 0.0499 & 0.0510 & 96.0 & $-$0.2521 & 0.3237 & 0.0511 & 0.0565 & 93.2\\
    MR$q'^2$ & \textbf{0.5721} & \textbf{0.5452} & \textbf{0.0467} & \textbf{0.0489} & \textbf{79.6} & \textbf{0.3908} & \textbf{0.3806} & \textbf{0.0478} & \textbf{0.0539} & \textbf{90.8}\\
    MR$p'^1p'^2$ & $-$0.0331 & 0.2607 & 0.0510 & 0.0524 & 95.0 & $-$0.1766 & 0.3065 & 0.0525 & 0.0580 & 95.0\\
    MR$p'^1q'^1$ & $-$0.0836 & 0.2573 & 0.0501 & 0.0512 & 94.4 & $-$0.2199 & 0.3181 & 0.0520 & 0.0568 & 93.8\\
    MR$p'^1q'^2$ & $-$0.0365 & 0.2591 & 0.0508 & 0.0519 & 95.6 & $-$0.1791 & 0.3087 & 0.0526 & 0.0575 & 94.6\\
    MR$p'^2q'^1$ & $-$0.0810 & 0.2490 & 0.0493 & 0.0515 & 95.6 & $-$0.2300 & 0.3171 & 0.0515 & 0.0570 & 94.6\\
    MR$p'^2q'^2$ & \textbf{0.5818} & \textbf{0.5427} & \textbf{0.0452} & \textbf{0.0487} & \textbf{78.0} & \textbf{0.4189} & \textbf{0.3943} & \textbf{0.0468} & \textbf{0.0537} & \textbf{90.8}\\
    MR$q'^1q'^2$ & $-$0.0894 & 0.2864 & 0.0528 & 0.0545 & 96.4 & $-$0.2467 & 0.3419 & 0.0531 & 0.0597 & 94.6\\
    MR$p'^1p'^2q'^1$ & $-$0.0826 & 0.2569 & 0.0501 & 0.0513 & 95.0 & $-$0.2181 & 0.3144 & 0.0517 & 0.0569 & 94.0\\
    MR$p'^1p'^2q'^2$ & $-$0.0235 & 0.2540 & 0.0504 & 0.0520 & 95.4 & $-$0.1680 & 0.3028 & 0.0524 & 0.0575 & 94.8\\
    MR$p'^1q'^1q'^2$ & $-$0.0831 & 0.2741 & 0.0517 & 0.0533 & 95.2 & $-$0.2195 & 0.3328 & 0.0534 & 0.0586 & 94.4\\
    MR$p'^2q'^1q'^2$ & $-$0.0781 & 0.3008 & 0.0543 & 0.0569 & 96.4 & $-$0.2228 & 0.3573 & 0.0555 & 0.0622 & 96.0\\
    MR$p'^1p'^2q'^1q'^2$ & $-$0.0792 & 0.2799 & 0.0524 & 0.0557 & 96.2 & $-$0.2196 & 0.3354 & 0.0536 & 0.0606 & 95.8\\
    \hline
\end{tabular*}
\vspace{1ex}
{\footnotesize
	\noindent\textsuperscript{a} Bias: mean bias ($\times 10$); MSE: mean squared error ($\times 10^2$); SD: standard deviation; $\overline{\text{SE}}$: average estimated standard error; CR: 95\% normality-based confidence interval coverage rate (\%).\\[0.5ex]
	\textsuperscript{b} Values in bold indicate results where all component working models are misspecified.
}
\end{table}

Additionally, Tables S1-S6 also compare several confidence interval construction methods, including the percentile and pivotal bootstrap methods. We observed that the coverage rates from these alternative methods were highly consistent with those from the normality-based approach, with all valid estimators yielding coverage rates close to the 95\% nominal level. The strong agreement between the computationally efficient normality-based CI and these more computationally intensive methods—combined with the stable performance across all censoring rates (0\% to 25\%)—provides strong empirical support for our approach. This verifies that the asymptotic normality of our estimator provides an accurate approximation in finite samples, confirming that the proposed normality-based confidence interval is reliable, valid, and practical for this application.

To further corroborate our findings beyond the primary settings, we conducted extensive supplementary simulations utilizing other data-generating processes, which are detailed in Section 1 of the Supplementary material. Specifically, Tables S7–S10 report the performance of the proposed estimator at a sample size of $n=500$ under an extended range of censoring proportions (0\%, 10\%, 25\%, and 50\%), while Tables S11–S14 present the corresponding results for a smaller cohort of $n=200$. Consistent with our main analysis, each table provides an exhaustive evaluation across all possible cross-combinations of the six candidate working models ($\{p^1, p^2, p^3, q^1, q^2, q^3\}$) and documents empirical coverage rates derived from the normality-based, percentile-based, and pivotal-based bootstrap methods. This pattern of reliable performance is consistently observed across all evaluated finite sample sizes and censoring scenarios. Collectively, the stability of these findings demonstrates that the proposed estimator is highly robust—not only to model misspecification, but also to varying degrees of right-censoring, fluctuations in sample capacity, and complex underlying data structures.

In summary, the simulation study confirms the effectiveness of the proposed estimator for competing risks data, demonstrating that it produces consistent estimates even if only one of the component PS or OR models is correctly specified.

\subsection{Real Data Analysis}
To illustrate the application of our proposed multiply robust method, we analyzed right heart catheterization (RHC) data. The RHC data were collected from an observation study on a RHC procedure with the aim to evaluate the effectiveness and safety of the RHC procedure for critically ill patients~\cite{connor1996}. The RHC data set contains information on 5735 critically ill patients who were considered for a RHC procedure upon their admission to a hospital. Among them, 2184 patients underwent the RHC procedure, while the remaining 3551 did not.

Given its observational nature and rich data, the RHC cohort has been a substrate for numerous methodological analyses. Previous analyses of these data have examined different outcomes and used various approaches. Several studies focused on 30-day survival~\citep{connor1996,tan2006distributional,li2018balancing,zhou2022addressing,bonvini2022sensitivity,cui2024selective}, while others analyzed length-of-stay by combining discharge and in-hospital death into a composite endpoint~\citep{mao2020flexible,keele2021comparing}. Vakulenko-Lagun et al.~\cite{vakulenko2023causalcmprsk} treated discharge and in-hospital death as competing events and estimated the effect of RHC on time-to-discharge and time-to-in-hospital-death; their results indicated a relatively higher early hazard of in-hospital death under RHC that diminished over time, becoming non-significant between approximately 50 and 150 days and later showing a lower death rate for the RHC group.

\begin{figure}[htbp]
	\centering
	\includegraphics[width=0.9\textwidth]{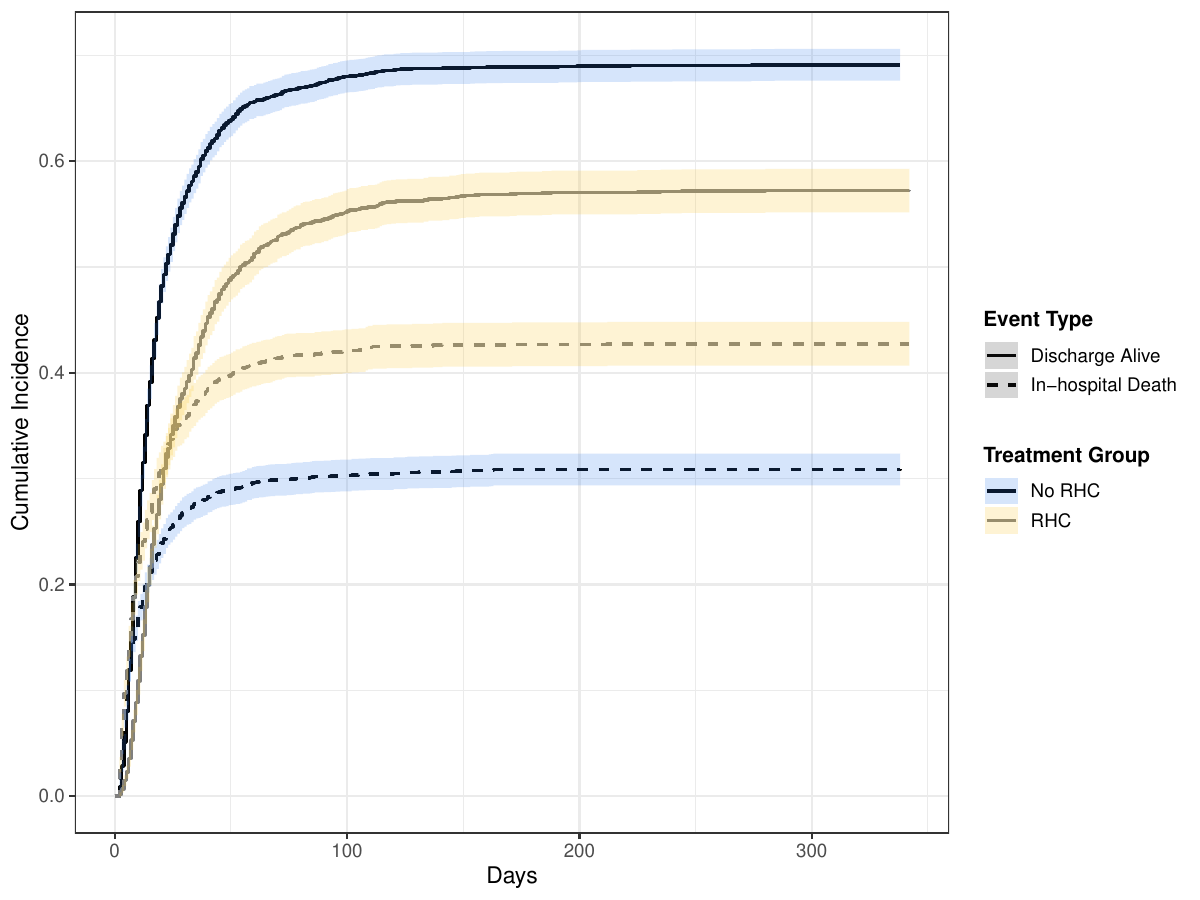}
	\caption{The cumulative incidence of two types of events terminated the hospitalization by treatments.}
	\label{fig:ci_rhc}
\end{figure}

In our analysis, we adopted this competing risks framework, defining discharge alive (competing event $1$) and in-hospital death (competing event $2$) as the two competing events. In the RHC data, 3704 patients were discharged alive, while 2030 patients died during their hospital stay. In addition, one observation with a missing date of discharge but a known date of death was removed from the study. We utilized the 65 baseline covariates included in the dataset, summarized in Table S15, to construct the propensity score and outcome regression models. Figure \ref{fig:ci_rhc} shows the estimated cumulative incidence curves for both event types, stratified by RHC treatment status, providing a preliminary visualization of the competing risks dynamics.

To specify the candidate models required for our multiply robust estimator, we first fitted a primary set of nuisance models. We used stepwise logistic regression on the treatment variable (\textit{swang1}) to fit the PS model, and stepwise Cox regression on survival time and status to fit the OR models. These fitted models served as the primary set of candidate models for the subsequent multiply robust analysis. The covariates selected by this procedure for the PS and OR models are detailed below:
\begin{enumerate}
	\item Model $p^1$ includes \textit{aps1, card, pafi1, resp1, paco21, dnr1, meanbp1, resp, neuro, hrt1, transhx, wtkilo1, ca, ninsclas, seps, hema1, liverhx, dementhx, hema, pot1, ph1, psychhx, trauma, gastr, chfhx, sod1, renal, edu, bili1, gibledhx, cardiohx, surv2md1, scoma1, crea1, renalhx, alb1}, and \textit{age}.
	\item Model $q^1$ includes \textit{surv2md1, swang1, ca, age, hema1, hrt1, alb1, dnr1, temp1, transhx, bili1, pafi1, neuro, chrpulhx, sod1, ninsclas, das2d3pc, chfhx, gibledhx, meta, resp1, wblc1, amihx, crea1, renalhx, income, gastr, resp, liverhx}, and \textit{seps}.
\end{enumerate}
Next, to construct a set of deliberately misspecified models for the multiply robust analysis, we performed a univariable screening of all 65 covariates. We assessed the association of each covariate with the treatment (RHC procedure) using univariable logistic regression, and with the survival outcome using univariate Cox regression. Based on the screening, we formulated these misspecified models (e.g., $p^2$ and $q^2$) by intentionally including the 10 covariates that exhibited the least statistical significance in their respective univariate analyses. The covariates selected for these deliberately misspecified PS and OR models are detailed below:
\begin{enumerate}
	\item Model $p^2$ includes \textit{hema, age, liverhx, ninsclas, race, renalhx, meta, ortho, pot1}, and \textit{temp1}.
	\item Model $q^2$ includes \textit{sex, pot1, ninsclas, malighx, das2d3pc, immunhx, ph1, gibledhx, wtkilo1}, and \textit{ortho}.
\end{enumerate}

Using these two sets of candidate models—the data-fitted set ($p^1, q^1$) and the deliberately misspecified set ($p^2, q^2$)—we proceeded to estimate the causal effect of RHC. We compared the performance of several estimators:
(i) standard IPW, using only $p^1$ or $p^2$;
(ii) standard outcome regression, using only $q^1$ or $q^2$; and
(iii) our proposed multiply robust estimators, which combine multiple PS and OR models.
The parameter of interest was the cause-specific CIF difference for discharge alive at several key time points: $t = 10, 20, 30, \text{and } 40$ days.

\begin{figure}[htbp]
	\centering
	\includegraphics[width=\textwidth]{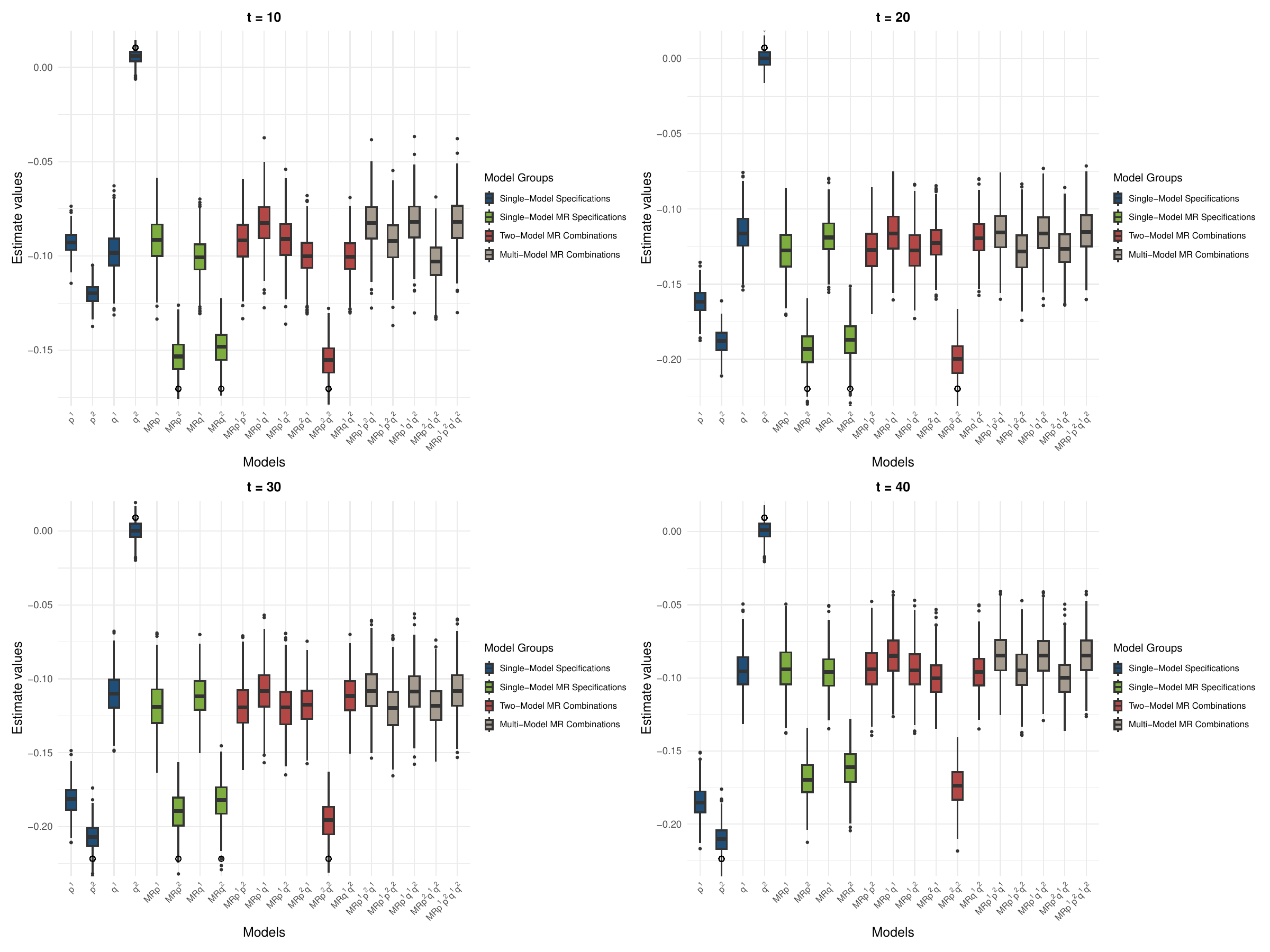}
	\caption{The RHC data analysis: estimate $\hat{\Delta}_1(t)$ at $t=10,20,30,40$ days using different methods.}\label{fig:rhc_results}
\end{figure}

From Figure \ref{fig:rhc_results} and Table S16, most of the estimators indicate that within 50 days after receiving the treatment, the RHC procedure exerted a negative effect on the pre-discharge survival status of the patients. However, there are some differences among the values of different estimators. As shown, the estimates from the standard, single-model estimators are highly sensitive to model specification. For example, the CIF difference estimated by the IPW($p^2$) model was substantially lower than that from IPW($p^1$). Conversely, the estimate from the OR($q^2$) model was significantly higher than that from OR($q^1$). This instability is also observed in multiply robust estimators that only included the misspecified models (e.g., MR$p^2q^2$), yielding results divergent from the other multiply robust combinations. In contrast, the full multiply robust estimator (MR$p^1p^2q^1q^2$), which incorporates all candidate models, demonstrates clear robustness to the inclusion of the misspecified models. This full multiply robust model estimates that the differences in the cumulative incidence of discharge alive at 10, 20, 30, and 40 days are -0.0819 (95\% CI: -0.0831, -0.0808), -0.1148 (95\% CI: -0.1161, -0.1134), -0.1081 (95\% CI: -0.1094, -0.1067), and -0.0842 (95\% CI: -0.0856, -0.0829), respectively. These results suggest that, for critically ill patients, the probability of being discharged alive was 8.2, 11.5, 10.8, and 8.4 percentage points lower in the RHC treatment group than in the control group at the respective time points. All effects were statistically significant at the 95\% level.

Furthermore, to ensure that our clinical findings are not strictly sensitive to the initial variable selection strategy, we conducted an extensive sensitivity analysis on the RHC dataset. Specifically, we employed LASSO regularization to empirically specify the models for both the propensity score and the outcome regression. This alternative, data-driven approach yielded highly consistent estimates of the CIF difference across the evaluated time points ($t=10, 20, 30,$ and $40$ days), thereby strongly corroborating the practical robustness of our proposed estimator. The detailed experimental setup and comprehensive results are deferred to Section 2 of the Supplementary material. Within this section, Figure S1 provides a visual comparison of the estimation results across all evaluated estimators. To supplement this figure, Table S17 details the complete numerical estimates along with the specific subset of covariates selected by LASSO.

\section{Discussion}
In this article, we have developed a novel multiply robust framework for estimating causal treatment effects in the presence of right censoring and competing events at time $t$. Simulation studies confirmed that a critical advantage of this methodology is its ability to address the limitations of existing doubly robust estimators, which typically exhibit considerable bias and poor coverage under dual misspecification. In contrast, our proposed estimator ensures consistency and superior finite-sample performance provided that the union of candidate models contains at least one correctly specified model. 
Crucially, extensive numerical evaluations and real-data applications demonstrate that our framework maintains strict robustness and high estimation accuracy across varying sample sizes, diverse underlying data complexities, and alternative variable selection strategies (e.g., stepwise regression versus LASSO). This stability persists even under substantial censoring proportions.
While our primary focus was on the cause-specific CIF difference, the proposed framework is flexible and readily generalizable to other key time-to-event estimands, such as the marginal cumulative incidence curve, restricted mean time lost, and survival quantiles.

While the proposed multiply robust framework addresses key challenges in causal estimation, several limitations warrant further discussion. First, the pseudo-value approach used in our method hinges on the assumption of independent censoring~\citep{andersen2010pseudo}. As a result, our current framework applies only under noninformative censoring, and the resulting estimator may be sensitive to dependent censoring. Although extensions of the pseudo-value technique for dependent censoring have been proposed~\citep{binder2014pseudo, overgaard2019pseudo}, incorporating them into a multiply robust framework entails nontrivial theoretical development and remains an important direction for future work. Second, like most causal inference methods for observational studies, our estimator relies on the assumption of no unmeasured confounding. Because this assumption is fundamentally untestable, potential violations may introduce bias, highlighting the need to combine the proposed approach with sensitivity analysis techniques~\citep{shi2020multiply, baiocchi2014instrumental} in future research. 

Moreover, we found that confidence interval coverage rates deteriorate as the censoring proportion increases. This degradation is expected, since heavy censoring reduces the available information and statistical power, thereby impairing the precision and reliability of inference. Finally, our confidence interval estimation relies on bootstrap-based approaches. While effective, the observed variation across different bootstrap methods highlights the need for future research to derive a more broadly applicable asymptotic variance formula, which would facilitate more efficient inference.

Looking forward, our proposed multiply robust framework holds significant potential for broader application in complex biomedical settings. For instance, in large-scale observational studies utilizing electronic health records, our method could be pivotal for accurately evaluating the comparative effectiveness of novel cancer therapies where death constitutes a significant competing risk and treatment assignment is non-randomized. Beyond these applications, a natural extension involves adapting the methodology to handle interval-censored data, which frequently arise when event times are observed only periodically, as well as generalizing the framework to multi-state processes to capture transitions between intermediate health states. Complementing these structural expansions, future research could also enhance the estimation procedure by data-adaptive machine learning techniques, such as ensemble methods or cross-fitting strategies. Such integration would offer robust protection against model misspecification, especially in high-dimensional settings where parametric assumptions are difficult to verify.

\section{Conclusion}\label{sec13}

Under standard causal assumptions, including consistency, positivity, and conditional exchangeability, we propose a multiply robust framework for estimating cause-specific CIF differences in the presence of right censoring and competing risks. By integrating the pseudo-value approach with multiple candidate propensity score and outcome regression models, the proposed estimator substantially reduces sensitivity to model misspecification compared with conventional doubly robust methods. The key strength of the proposed framework lies in its multiply robust property: consistency is guaranteed provided that at least one candidate propensity score or outcome regression model is correctly specified. Extensive simulation studies demonstrate that this property leads to improved finite-sample performance, with negligible bias and near-nominal coverage probabilities across a wide range of censoring rates and substantial model misspecification. Moreover, the pseudo-value formulation avoids reliance on proportional hazards assumptions and facilitates direct modeling of marginal causal estimands. An application to the RHC dataset illustrates the practical utility of the proposed method. Beyond the cause-specific CIF difference, the framework is flexible and can be extended to other time-to-event estimands, such as restricted mean time lost and survival quantiles. Overall, this methodology provides a robust and practical tool for causal inference in observational studies with competing risks and complex confounding structures.

\section*{Supplementary information}

The supplementary file accompanying this article (Supplement.pdf) provides additional numerical results and data analysis to support the main findings. Specifically, it contains:
\begin{enumerate}
	\item Simulation Results: Table S1 through Table S6, which present the complete simulation results for scenarios I-III under various conditions (with and without data censoring), including Mean Bias, Mean Squared Error (MSE), Standard Deviation (SD), and Confidence Interval Coverage Rates (CR). Tables S7-S14 detail the finite-sample performance of various model combinations across different sample sizes ($n=500$ and $n=200$) and varying censoring proportions (0\%, 10\%, 25\%, and 50\%) in supplementary scenarios (Scenarios IV–VI).
	\item Real Data Analysis: The baseline characteristics table for the Right Heart Catheterization (RHC) study, providing a comprehensive summary of the study population. Detailed estimation results of the CIF difference for the RHC data at $t=10, 20, 30,$ and $40$ days, comparing the performance of different methods used in the study.
	Furthermore, it details an alternative sensitivity analysis utilizing LASSO for variable selection (Tables S17 and Figure S1) to estimate the CIF differences at $t=10, 20, 30,$ and $40$ days.
\end{enumerate}

\section*{Appendix}
\begin{appendices}
	
	\section{The estimation process of $\hat{F}_k^{(-i)}$ using the Aalen–Johansen estimator}\label{app:pseudo_CIF_est}
	
	We define the counting processes $\tilde{N}_{ik}(t)=\mathbb{I}(\tilde{T}_i\le t, \tilde{\epsilon_i}=k)$ and $Y_i(t)=\mathbb{I}(\tilde{T}_i > t)$. The Kaplan-Meier estimate of survival function of the censoring time $\mathcal{G}(t)$ could be expressed as
	\begin{equation*}
		\hat{\mathcal{G}}(t)=\prod_{s=0}^t\left[1-\frac{\widehat{H_0}(ds)}{n^{-1}\sum_{i=1}^n Y_i(s-)}\right],
	\end{equation*}
	where $\widehat{H_0}(t) = n^{-1}\sum_{i=1}^n \mathbb{I}(\tilde{T}_i \le t, \tilde{\epsilon}_i=0)$. The inverse of the probability of censoring weighted (IPCW) estimator of the cause-specific CIF $F_k(t)$ is given by $$\hat{F}_k(t)=\frac{1}{n}\sum_{i=1}^n \int_0^t \frac{\tilde{N}_{ik}(ds)}{\hat{\mathcal{G}}(s-)}.$$
	Then the Aalen-Johansen estimate of $F_k(t)$ for $\boldsymbol{D}^{(-i)}$ is given as: 
	$$\hat{F}_k^{(-i)}(t)=\frac{1}{(n-1)}\sum_{j\neq i}\int_0^t\frac{\tilde{N}_{jk}(ds)}{\hat{\mathcal{G}}^{(-i)}(s-)}\;, \quad i = 1, \ldots, n\;,$$
	where $\hat{\mathcal{G}}^{(-i)}$ is the Kaplan-Meier estimate for $\mathcal{G}(t)$ based on $\boldsymbol{D}^{(-i)}$.

\end{appendices}

\bibliographystyle{unsrt}  
\bibliography{competing_refer}

\end{document}